# Natural van der Waals canalization lens for non-destructive nanoelectronic circuit imaging and inspection

Qingdong Ou[1,2]*†, Shuwen Xue[3]†, Weiliang Ma[4], Jiong Yang[5], Guangyuan Si[6], Lu Liu[7], Gang Zhong[1], Jingying Liu[1,2], Zongyuan Xie[7], Ying Xiao[7], Kourosh Kalantar-Zadeh[8], Xiang Qi[9], Peining Li[10], Zhigao Dai[7,11]*, Huanyang Chen[3]*, Qiaoliang Bao[9,12,13]*

[1]Macao Institute of Materials Science and Engineering (MIMSE), Faculty of Innovation Engineering, Macau University of Science and Technology, Taipa, Macao 999078, China.

[2]Department of Materials Science and Engineering, Monash University, Clayton, Victoria 3800, Australia.

[3]Department of Physics, Xiamen University, Xiamen 361005, China.

[4]National Engineering Research Center of Electromagnetic Radiation Control Materials, Key Laboratory of Multi-spectral Absorbing Materials and Structures of Ministry of Education, University of Electronic Science and Technology of China, Chengdu, 610054, China

[5]School of Optoelectronic Science and Engineering & Collaborative Innovation Center of Suzhou Nano Science and Technology, Key Lab of Advanced Optical Manufacturing Technologies of Jiangsu Province & Key Lab of Modern Optical Technologies of Education Ministry of China, Soochow University, Suzhou 215006, China.

[6]Melbourne Centre for Nanofabrication, Victorian Node of the Australian National Fabrication Facility, Clayton, Victoria 3168, Australia.

[7]Engineering Research Center of Nano-Geomaterials of Ministry of Education, Faculty of Materials Science and Chemistry, China University of Geosciences, Wuhan 430074, China.

[8]School of Chemical and Biomolecular Engineering, The University of Sydney, Sydney, New South Wales, Australia.

[9]Hunan Key Laboratory for Micro-Nano Energy Materials and Devices, School of Physics and Optoelectronic, Xiangtan University, Hunan, 411105 China

[10]Wuhan National Laboratory for Optoelectronics and School of Optical and Electronic Information, Huazhong University of Science and Technology, Wuhan 430074, China.

[11]Shenzhen Research Institute China University of Geosciences, Shenzhen 518063, China.

[12]Institute of Energy Materials Science (IEMS), University of Shanghai for Science and Technology, Shanghai 200093, China.

[13]Nanjing kLIGHT Laser Technology Co., Ltd., Nanjing, Jiangsu 210032, China.

†These authors contributed equally to this work.

*Correspondence to: qdou@must.edu.mo (Q.O.); daizhigao@cug.edu.cn (Z.D.); kenyon@xmu.edu.cn (H.C.); qiaoliang.bao@usst.edu.cn (Q.B.)

**Abstract**

**Optical inspection has long served as a cornerstone non-destructive method in semiconductor wafer manufacturing, particularly for surface and defect analysis. However, conventional techniques such as bright-field and dark-field scattering optics face significant limitations, including insufficient resolution and the inability to penetrate and detect buried structures. Atomic force microscopy (AFM), while offering higher resolution and precise surface characterization, is constrained by slow speed, limited to surface-level imaging, and incapable of resolving subsurface features. Here, we propose an approach that integrates the strengths of dark-field scattering optics and AFM by leveraging a van der Waals (vdW) canalization lens based on natural biaxial α-MoO$_3$ crystals. This method enables ultrahigh-resolution subwavelength imaging with the ability to visualize both surface and buried structures, achieving a spatial resolution of 15 nm and grating pitch detection down to 100 nm. The underlying mechanism relies on the unique anisotropic properties of α-MoO$_3$, where its atomic-scale unit cells and biaxial symmetry facilitate the diffraction-free propagation of both evanescent and propagating waves via a flat-band canalization regime. Unlike metamaterial-based superlenses and hyperlenses, which suffer from high plasmonic losses, fabrication imperfections, and uniaxial constraints, α-MoO$_3$ provides robust and aberration-free imaging in multiple directions. We successfully applied this approach to achieve high-resolution inspection of buried nanoscale electronic circuits, offering unprecedented capabilities essential for next-generation semiconductor manufacturing.**

Since the advent of integrated circuits, optical inspection technology has been a cornerstone of semiconductor wafer manufacturing, providing a fast and non-destructive approach to defect detection and quality control, including bright-field and dark-field scattering techniques, plays a key role in ensuring production yields, but it faces fundamental limitations when nanoscale precision is required[1-4]. For instance, bright-field detection offers high resolution but struggles with low defect contrast[6], while dark-field methods can effectively highlight surface defects but fail to detect subsurface features[71,8]. Atomic force microscopy partially overcomes the resolution barrier, offering nanoscale precision for surface measurements. However, AFM is restricted to surface imaging, and incapable of penetrating layers to reveal buried structures, greatly limiting its applicability for inspecting advanced multi-layered semiconductor devices.[4,6] As the semiconductor industry approaches the physical limits of Moore's law, the increasing complexity of AI and quantum chips has driven the need for novel inspection technologies capable of probing multi-dimensional structures, including deeply buried features, while maintaining high resolution and throughput.

Here, we propose an innovative method that integrates dark-field scattering optics and atomic force microscopy (AFM) with a van der Waals (vdW) canalization lens based on α-$MoO_3$. This hybrid approach enables ultrahigh-resolution subwavelength imaging, capable of resolving deeply buried features and overcoming the inherent limitations of conventional optical and AFM methods. Specifically, our system achieves a remarkable resolution of 15 nm, surpassing the capabilities of traditional optical detection limited by diffraction and AFM's surface-bound imaging. While scanning near-field optical microscopy (SNOM) has previously demonstrated its potential to achieve nanoscale imaging through near-field interactions, its effectiveness has often been restricted by the properties of uniaxial materials such as hexagonal boron nitride (hBN). Subwavelength imaging using hBN and other hyperbolic metamaterials, which rely on gapped hyperbolic dispersions, suffers from the image magnification effect and momentum mismatch, leading to losses and aberrations in the transmitted geometric information (Fig. 1a). Moreover, the uniaxial nature of these materials, characterized by identical in-plane permittivity components, limits imaging to a single direction (Fig. 1c), making multidirectional imaging unattainable. Despite their successes, these limitations highlight the need for more advanced materials and methods to unlock the full potential of near-field imaging.

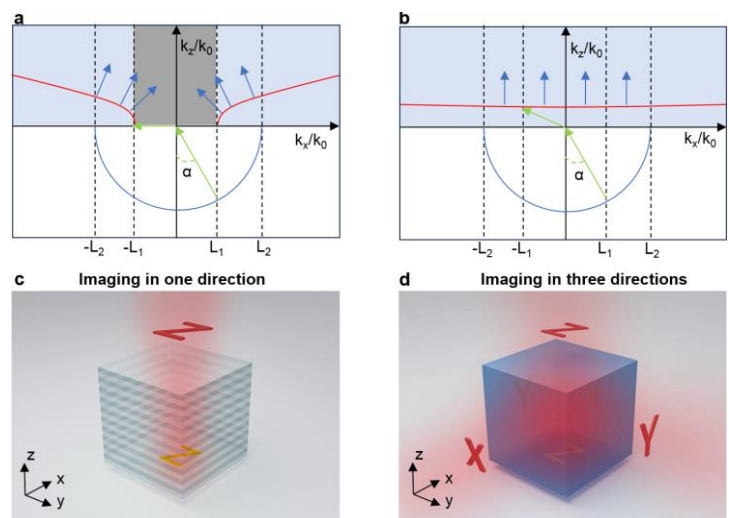

**Fig. 1. Schematic diagrams of subwavelength imaging enabled by hyperbolic or flattened photonic dispersions.** (a) Imaging with a gapped hyperbolic dispersion in momentum space,

which leads to unavoidable aberration. The upper plane shows photonic dispersions of the imaging lens (red solid lines in blue shaded region), and the lower plane (blue solid lines) shows dispersions of the objects. The black dashed vertical lines are the equal-$k_x$ lines, which intersect with the dispersions at positions $L_1$ (-$L_1$) and $L_2$ (-$L_2$), respectively. The gray shaded region suggests information lost due to the dispersion gap in the lens. The green arrows denote the wave vectors of the incident and transmitted waves. The blue arrows denote the Poynting vectors. $k_0$: wave vector in free space. α: incident angle. (b) Imaging with flat gapless dispersion. Blue circle refers to the dispersion of air, while the gapped and gapless dispersions (red lines) belong to different imaging media. The arrows indicate the transmission directions of geometric information. (c) Imaging in one direction is allowed by using uniaxial media such as layered hyperbolic metamaterials and hBN crystal. (d) Imaging in three directions can be achieved by biaxial media such as α-MoO$_3$ crystal proposed in this work. X, Y and Z: subwavelength objects and images across respective lens planes.

In contrast, the natural biaxial vdW crystal α-MoO$_3$ overcomes these challenges by offering unique anisotropic properties with distinct permittivity components along orthogonal crystal axes. These properties allow precise control of light propagation, enabling the diffraction-free transmission of all wavevector components through its gapless flat dispersions (Fig. 1b). As a result, highly collimated and canalized phonon polariton (PhP) modes propagate with constant longitudinal momentum, carrying geometric information without loss or aberration. Unlike systems based on artificial superlenses or hyperlenses, which are limited by fabrication imperfections, plasmonic losses, and large meta-atom sizes, α-MoO$_3$ achieves near-lossless subwavelength imaging through its naturally occurring anisotropy. More importantly, compared to conventional optical and AFM methods, α-MoO$_3$-based imaging enables robust visualization of both surface and buried structures with exceptional resolution and stability. Furthermore, the canalization mechanism ensures consistent imaging performance even when the object-to-imaging-plane distance varies (Supplementary Note 2 and Supplementary Fig. 2), establishing α-MoO$_3$ as a transformative solution for next-generation multidimensional, high-resolution inspection technologies.

From Figure 2a, the α-MoO$_3$ imaging device consists of an α-MoO$_3$ slab positioned on top of subdiffractional objects made of Au nanopatterns. Imaging can be achieved through the flat-band canalization of PhPs in the mid-infrared range. In particular, the permittivity values of orthorhombic α-MoO₃ undergo three significant transitions, shifting from positive to negative along the orthogonal axes over the range (Fig. 2b and the inset), while hexagonal hBN only has two such transitions[23]. Notably, this unique feature of α-MoO₃ leads to subwavelength imaging in three directions (Fig. 1d), respectively, as revealed by theoretical modelling in Supplementary Note 1. This imaging character is barely attainable in previously reported plasmonic or metamaterial lenses due to higher structural and optical symmetries. Furthermore, these topological transitions induce the flat bands of PhPs in α-MoO₃, which is nontrivial for super-resolution subwavelength imaging. Similar topological transitions have been recently observed in twisted bilayers of α-MoO₃ [42-45] and graphene/α-MoO₃ heterostructures[46-48] with in-plane canalization and diffractionless propagation of PhPs; however, the application of this mechanism for imaging remains largely unexplored.

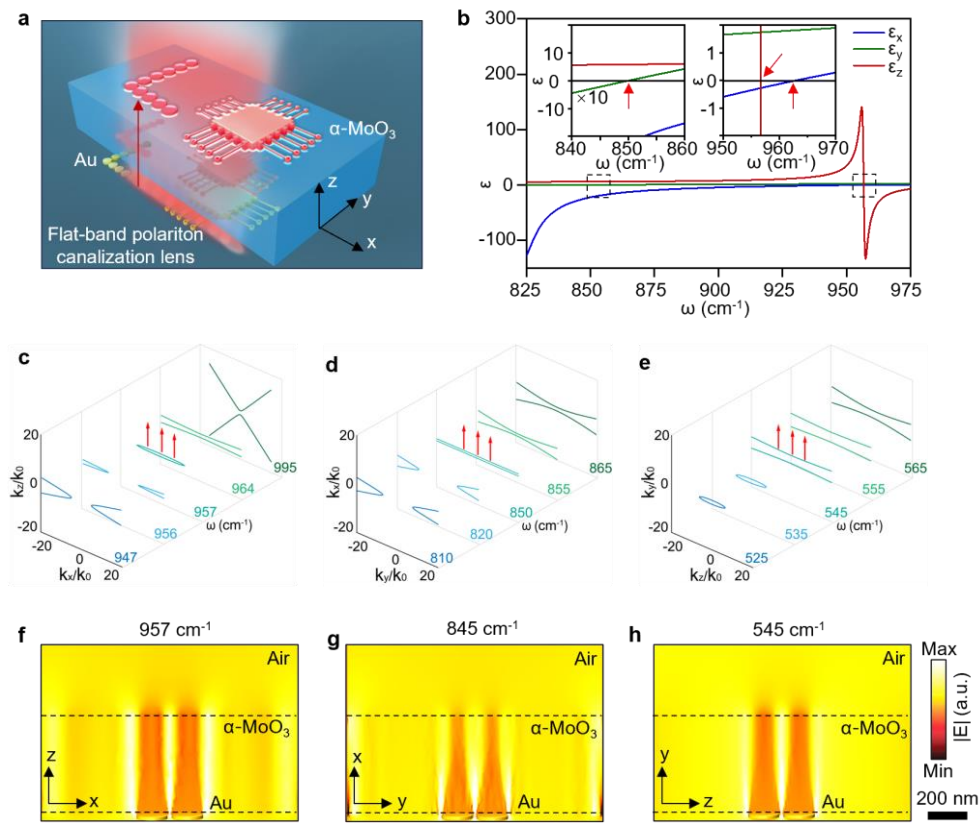

**Fig. 2. Working principle of ultrahigh-resolution imaging at multiple canalization frequencies with α-MoO₃.** (**a**) Schematic of the subwavelength imaging device comprised of a flat van der Waals crystal of α-MoO₃. A global Cartesian coordinate is defined such that the x, y and z axes are along the [100], [001] and [010] crystal directions of α-MoO₃. (**b**) Real parts of permittivity tensor components of α-MoO₃. Insets show the frequency ranges with positive to negative permittivity transitions in three crystal directions. (**c**) The evolution of isofrequency contours (IFCs) of α-MoO$_3$ across the topological transition frequency in the zx plane. (**d**) The IFCs in the xy plane. (**e**) The IFCs in the yz plane. The flattened IFCs in c-e suggest the canalization regime of low-loss phonon polaritons, which is used to restore the images through α-MoO$_3$. The arrows on the ultraflat curve, that is, the Poynting vectors of polaritons with the almost same radial direction, indicate the diffractionless polariton propagation along certain directions. (**f-h**) Imaging in three directions at multiple frequencies with α-MoO$_3$. Simulated cross-section electric field distributions (|E|) of a 500 nm-thick α-MoO$_3$ lens show deep-subwavelength imaging in three directions respectively of two closely-packed Au objects with an interparticle gap of 15 nm.

Topological transitions and canalization regimes exist in all the three directions for a single layer α-MoO$_3$. As shown in Fig. 2c-e, the isofrequency contours (IFCs) in the z and x directions of α-MoO$_3$ evolve from hyperbolic to elliptical and then 90$^\circ$-rotated hyperbolic shapes, while the IFCs in the y direction directly transit from elliptical to hyperbolic. Near the topological transition frequencies, both the hyperbolic and elliptical IFCs are strongly flattened (for more details, see Supplementary Note 3 and Supplementary Figs. 3-5). Thus, extremely collimated and canalized PhP modes carrying the geometric information can propagate simultaneously with a constant longitudinal momentum and a fixed upward direction along the z direction of α-MoO$_3$, which will be utilized to restore the super-resolution images by α-MoO$_3$. This conceptual hypothesis has been confirmed by the numerically simulated field distributions of a 500 nm-thick α-MoO$_3$ lens (Fig. 2f-h); this theoretically enables subwavelength imaging in three directions. Moreover, the α-MoO$_3$ lens provides extra options with multiple canalization-based imaging frequencies from around 540 cm$^{-1}$ to 960 cm$^{-1}$ while only one imaging frequency is allowed for a conventional uniaxial lens.

To experimentally observe the topological transition and subwavelength images, we used a real-space optical technique to directly map the objects through scattering-type scanning near-field optical microscopy (s-SNOM). An oscillating tip under free-space illumination acts as a near-field probe as the surface of the α-MoO$_3$ imaging device is scanned. This technique allows the simultaneous recording of nanoscale-resolved near-field images with topography. Figure 3a-c shows the near-field images of 93 nm-thick α-MoO$_3$ on top of a 500 nm-diameter Au disk under light illumination of 897, 957 and 995 cm$^{-1}$, respectively. The wavefronts and dispersions of PhPs change from hyperbolic to elliptical shapes, revealing a topological transition (Supplementary Fig. 6). These in-plane elliptical wavefronts arise from the out-of-plane hyperbolic dispersion of α-MoO$_3$ in nature as shown in Fig. 2c and e. When the out-of-plane hyperbolicity reduces to nearly flat contours, both the elliptical and hyperbolic wavefronts disappear, restoring the images of the underlying objects at the surface of α-MoO$_3$ (Fig. 3b).

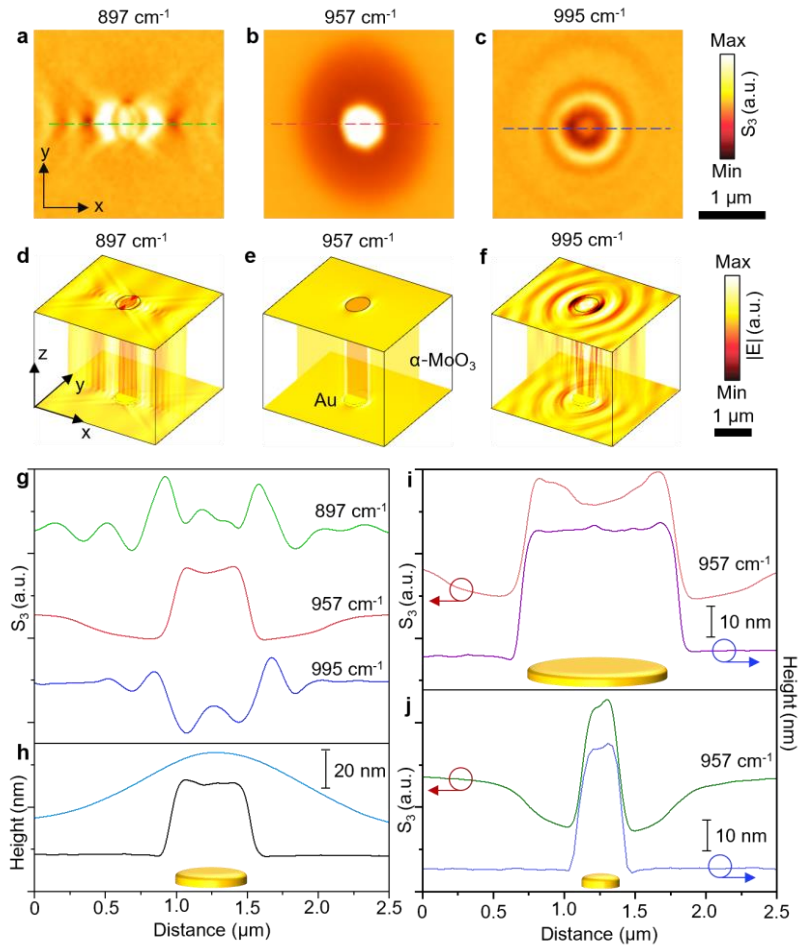

**Fig. 3 Topological transition of phonon polaritons in α-MoO$_3$.** (**a-c**) Experimentally measured near-field images on the top surface of 93 nm-thick α-MoO$_3$ sitting on an individual Au disk (diameter ~500 nm) at 897, 957 and 995 cm$^{-1}$, respectively. Scale bars for **a-c**, 1 μm. (**d-f**) Corresponding simulated electric field distributions (|E|), which are artificially restacked in three dimensions. Scale bars for **d-f**, 1 μm. (**g**) Line traces of near-field images along the x direction in **a-c**, respectively. (**h**) Line traces of AFM images of Au disk before (black line) and after (azure blue line) transfer of α-MoO$_3$ in Fig. 3a-c. Scale bar for height, 20 nm. (**i,j**) Line traces of near-field images and AFM images of the Au disks with diameter of 1000 nm (**i**)

and 250 nm (**j**), respectively. Scale bars for height, 10 nm. Insets show schematics of the corresponding Au disk in **h-j**.

To theoretically validate our experimental results, we performed three-dimensional full-wave simulations based on the finite element method. The above experimental observations are effectively verified by the numerically simulated images (Fig. 3d-f). The cross-section field distributions at the topological transition show that PhP modes propagate upwards with indistinguishable diffraction. This is reflected by extracting line profiles across the Au disk (Fig. 3g), with only one strong signal and a steep intensity rise/fall at 957 cm$^{-1}$. The near-field signal greatly resembles the topographic shape of the Au disk (Fig. 3h). The PhP canalization and hyperlensing effect are also verified in Au disks of different sizes (Fig. 3i and j and Supplementary Fig. 7). For example, the image signal and the topographic signal of a 250 nm object ($\lambda$/41.8) are almost the same considering the peak width. Note that the dark annular regions in Fig. 3b and the signal intensity drop in Fig. 3i and j are attributed to the spreading of suspension of $\alpha$-MoO$_3$ around the Au disks, indicating that polariton canalization is highly sensitive to the dielectric environment.

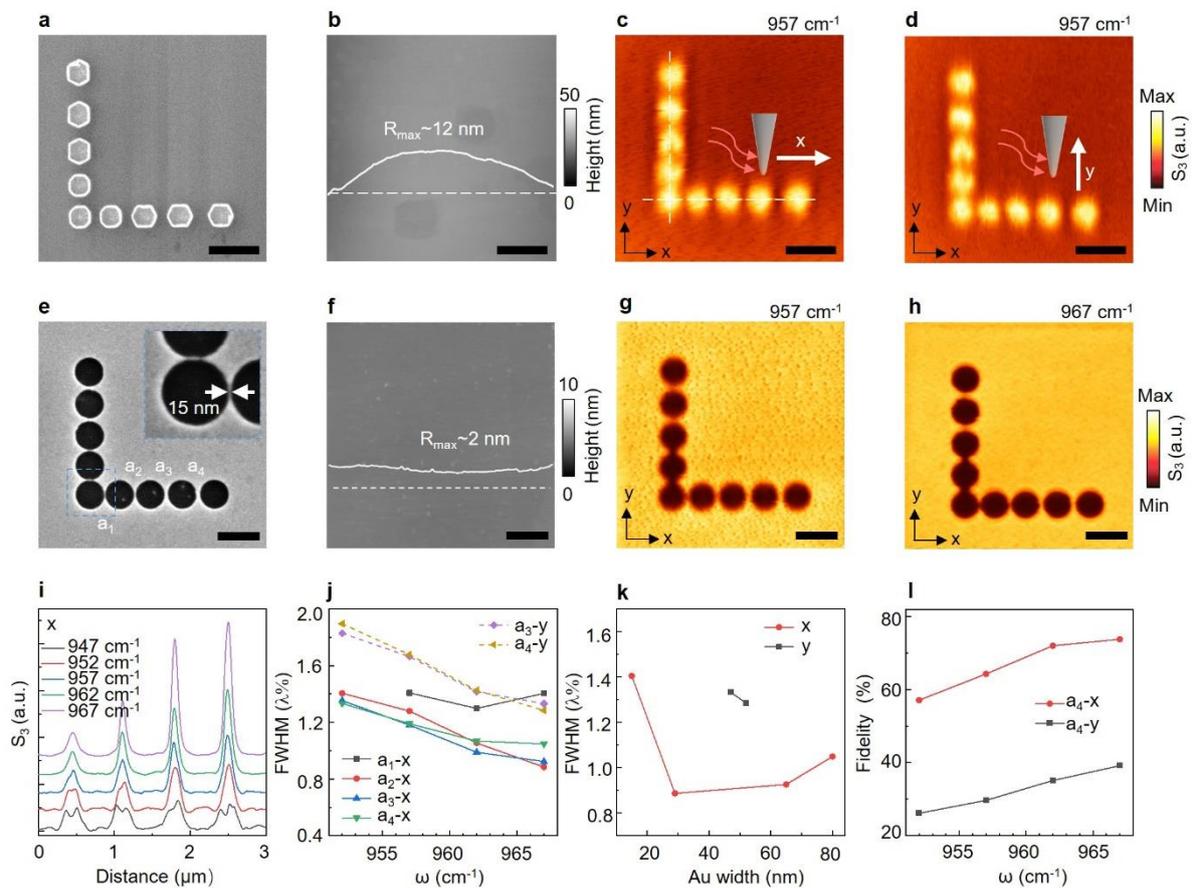

**Fig. 4 Ultrahigh-resolution α-MoO$_3$ imaging resolving deeply subdiffractional objects.** (**a**) SEM image of the L-shaped Au disk nanostructures on SiO$_2$/Si, with a narrowest gap of 68 nm and a thickness of 20 nm. (**b**) Surface topography of the 200 nm-thick α-MoO$_3$ lens, showing a hump of ~12 nm traced along the white dashed line. (**c, d**) Near-field images on the top surface of the α-MoO$_3$ lens at 957 cm$^{-1}$. The images were measured along different scanning directions. Scale bars for **a-d**, 500 nm. (**e**) SEM image of the L-shaped air-hole nanopatterns in the Au film. The widths of Au line between neighbouring holes along horizontal direction are a$_1$ = 15 nm, a$_2$ = 29 nm, a$_3$ = 65 nm and a$_4$ = 80 nm. Inset shows zoomed-in image of the extremely narrow Au line with a width of 15 nm. Scale bars for **e-h**, 1 μm. (**f**) Surface topography of the 230 nm-thick α-MoO$_3$ lens, showing a flat profile with only 2 nm fluctuation. (**g, h**) Near-field images measured on the top surface of α-MoO$_3$ lens at 957 cm$^{-1}$ and 967 cm$^{-1}$, respectively. (**i**) Line traces of near-field signals along the white dashed lines in **f** at different frequencies. The symbol 'x' represents the line traces extracted along the *x* direction. (**j**) Imaging performance by fitting the peak FWHM in the corresponding line traces. Note that a$_3$-y and a$_4$-y represent the widths of Au line between top three neighbouring holes in **e** along vertical direction. (**k**) Imaging performance calculated at 967 cm$^{-1}$ relative to the real object sizes. (**l**) Imaging fidelity defined by the ratio of real Au line width to the peak FWHM.

To explore the imaging limit of α-MoO$_3$, we designed and fabricated two sets of L-shaped objects with deep subwavelength sizes, as shown in Fig. 4. Two sets of samples are examined: reduced spacings down to 68 nm between two Au disks (Fig. 4a), and extremely narrow Au lines as small as 15 nm in width (Fig. 4e). α-MoO$_3$ exhibits ultrahigh imaging resolution at the frequency where the PhP canalization occurs. Although the Au nanostructures are completely masked by the α-MoO$_3$ layer, the device can effectively resolve the interparticle gaps and Au lines beyond the diffraction limit. Figure 4b shows the topography image of an as-made 200 nm-thick α-MoO$_3$ imaging device; here, the closely packed Au disks (Fig. 4a) are encased between a large-area α-MoO$_3$ crystal and the SiO$_2$/Si substrate, with a 12 nm hump at the hyperlens surface, which may affect the imaging resolution. The recorded images at 957 cm$^{-1}$ are displayed in Fig. 4c and d; underlying L-shaped disks with discernible interparticle spacings are observed. The imaging resolution is not influenced by the tip scanning directions, indicating that the intrinsic PhP modes account for the aberration-free imaging performance. Detailed analysis shows that a high resolution of λ/97 is achieved when illuminating the 68 nm spacing at 957 cm$^{-1}$ (for more details, see Supplementary Fig. 8).

More strikingly, α-MoO$_3$ exhibits the capability to resolve the objects with a size down to 15 nm; this achievement has not been attained in previously reported superlenses or hyperlenses. Notably, the 15 nm feature size has already exceeded the limitation of our near-field microscopy with an AFM tip radius of ~25 nm, which might compromise the intrinsic resolving capability of α-MoO$_3$. Extremely narrow Au lines (Fig. 4e) were fabricated by milling two holes in a Au film via ultrahigh-resolution focused ion beam (FIB) lithography. In this case, the topography of a 230-nm-thick α-MoO$_3$ device showed an ultraflat surface, without the characteristic hump as those for the disk objects, which could further minimize the external effects on imaging performance. Two images recorded at 957 cm$^{-1}$ and 967 cm$^{-1}$ are displayed in Fig. 4g and h, respectively. The entire Au nanopatterns were clearly recovered, as shown in Fig. 4e, with a relatively weak contrast for the line with a 15 nm width. Nonetheless, a decent peak corresponding to the 15-nm-wide line could be clearly observed in each linescan from 957 to 967 cm$^{-1}$ (Fig. 4i). By analysing the peak width at 967 cm$^{-1}$, we obtained a resolution of λ/71.2 for the 15-nm-wide line (Fig. 4j). Notably, the highest resolution up to λ/113 was achieved for the 29-nm-wide line (Fig. 4k). For the objects aligned along the *y* direction, a maximum resolution of λ/78 was achieved for the 52 nm line. More images and analysis are available in Supplementary Fig. 9. Notably, all resolutions achieved by α-MoO$_3$, regardless of the anisotropic resolving feature, were higher than those with hBN imaging.

In addition to high resolution, superior imaging characteristics of high fidelity induced by ultraflat polariton dispersions are also achievable. While resolution represents the capability of a lens to resolve objects, fidelity quantifies the quality of the high-resolution images, and this factor was often overlooked or unattainable in previously reported hyperlenses or superlenses. However, the α-MoO$_3$ imaging exhibits exceptional image fidelity, with a maximum value of 74% in the *x* direction at 967 cm$^{-1}$ and 40% in the *y* direction; this was calculated based on the ratio of the Au line width to the FWHM (Fig. 4l).

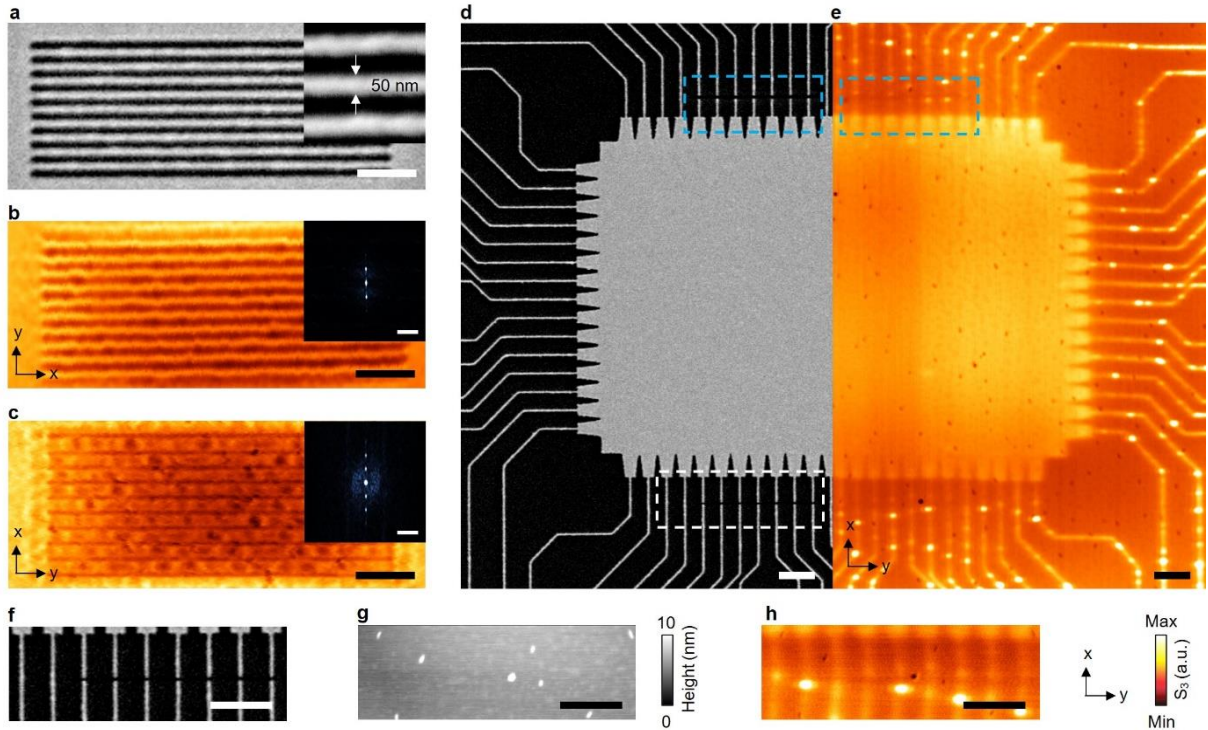

**Fig. 5 Application of α-MoO₃ imaging for inspection of buried nanoscale electronic circuit.** (**a**) SEM image of Au nanoribbon array. Inset shows the grating with a line width of 50 nm and a period of 100 nm. Scale bar, 500 nm. (**b,c**) Near-field images recorded on the top surface of 210 nm-thick α-MoO₃ at 967 cm$^{-1}$ with grating orientation along the y direction (**b**) and x direction (**c**). Scale bar, 500 nm. Insets in **b** and **c** show the corresponding FFT images. Scale bar, 150$k_0$. (**d**) FIB-SEM image of a "CPU" pattern with artificial cuts along Au traces (~40 nm line width). The top cut (blue dashed square) has a gap of ~60 nm while the bottom cut (white dashed square) has a gap of ~40 nm. The pattern was coated with an insulating SiO₂ layer before placing a 240 nm-thick α-MoO₃ flake. Scale bar, 1 μm. (**e**) Near-field image of the mirrored "CPU" pattern in **d**. Scale bar, 1 μm. (**f-h**) Zoom-in FIB-SEM image, surface topography and near-field image of the cuts of Au leads (see white dashed square region in **d**). Scale bar, 1 μm.

To fully examine the capability of imaging linear objects, we fabricated an imaging device with 210 nm-thick α-MoO₃ on top of a Au nanoribbon array. Figure 5a shows an FIB image of the Au nanoribbon array object with an average line width of 50 nm and a period of 100 nm written on Au film. The recorded field patterns at the image plane in Fig. 5b and c clearly show the profiles of the gratings captured by α-MoO₃ in both real space and momentum space. High spatial frequencies, at least the grating's third harmonics, are imaged by α-MoO₃ for both gratings oriented along the *x* and *y* directions (Supplementary Fig. 10). Thus, α-MoO$_3$ imaging can restore the details of the nanoscale linear patterns benefitting from diffraction-free canalization-based imaging properties, as discussed above.

Finally, we use α-MoO$_3$ imaging in an industrial application scenario in microelectronics. During the wafer fabrication process, the electronic devices are always protected by thin cap layers but need defect inspection and review; this is a critical part of outgoing wafer qualification. In this regard, a "CPU" pattern with conductive Au traces and a SiO$_2$ encapsulation layer resembling the processing unit of electronic chips (Fig. 5d) was fabricated. The measured image in Fig. 5e shows that the fine features of the underlying "CPU" object can be effectively reproduced with excellent fidelity. In particular, the bright spots in the recorded image (Fig. 5e) arise from localized polariton scattering, indicating residue or embedded particles along the Au traces. More importantly, the image shows the two locations of the artificial break or damage etched by high-resolution FIB (see two squared regions in Fig. 5d). The break with a deep subwavelength gap of 40 nm is efficiently resolved by a 240-nm-thick α-MoO$_3$ lens, as shown in the zoomed-in images of Fig. 5f-h. Due to this canalization-based imaging capability, α-MoO$_3$ is capable of quality control and break diagnosis for electronic circuits at the nanoscale.

We compared the resolving and imaging capability of α-MoO$_3$ canalization lens in our study to those of previously reported superlenses and hyperlenses (see Supplementary Fig. 11 and Table 1 for more details). Generally, natural hyperbolic crystal lenses have higher imaging resolutions than plasmonic and metamaterial lenses, which suffer from high losses and imperfect fabrication. For example, the SiC superlens can resolve air holes with a diameter of 540 nm milled in the Au film at 11 μm[13]; however, its resolution has recently been improved with synthetic waves of complex frequency[14], and a pair of Au holes with edge-to-edge separation of 40 nm was able to be resolved. In contrast, the monoisotopic h$^{11}$BN hyperlens is able to resolve an interparticle gap of 25 nm by using an image reconstruction algorithm[31]. In our study, α-MoO$_3$ is able to directly resolve an object as small as 15 nm, that is, a size of 0.00145λ; this resolution is among the best recorded to date.

The ultrahigh resolution of α-MoO$_3$ imaging is attributed to its inherent low loss and indefinitely large anisotropy. α-MoO$_3$ has an extremely low optical loss, which differs from the high plasmonic loss in metamaterials[23,49] and large phonon damping in hBN ($\gamma \sim 3$ and 7.3 cm$^{-1}$, respectively, for two Reststrahlen bands[23], while $\gamma \sim 1.5$ cm$^{-1}$ for α-MoO$_3$ in the out-of-plane direction[50]). More importantly, the extremely large anisotropy in α-MoO$_3$ enables precise diffraction control, resulting in flat-band polariton canalization with flattened gapless dispersions for ultrahigh-resolution canalization lens. Furthermore, these imaging channels exist in three directions. Different from most uniaxial hyperbolic metamaterials and hBN lenses restricting two-dimensional imaging, biaxial MoO$_3$ shows symmetry breaking of both in-plane and out-of-plane structural and optical properties, enabling subwavelength imaging in all the three directions. Due to limitations of our laser wavelength, we experimentally verified the imaging effect at one canalization region as an example.

In conclusion, we demonstrate ultrahigh-resolution deep-subwavelength canalization lens fabricated from a lithography-free biaxial vdW crystal; this α-MoO$_3$-based imaging is distinctive in comparison to all previously reported subwavelength lenses. We anticipate further improved resolution by using synthetic excitation waves of complex frequency[14] and tuneable imaging by controlling PhPs through intercalation[51,52]. The imaging frequencies of the vdW crystals can also extend from the terahertz to visible range by leveraging the vast library of naturally hyperbolic crystals[25,26]. Moreover, near-field microscopy based on vdW crystals can be utilized as an alternative way to resolve and image extremely small subwavelength objects that are not directly accessible by a near-field probe. The results from our study can be used to advance imaging resolution into a new stage extending beyond the diffraction limit and may facilitate opportunities for three-dimensional broadband optical imaging of nanostructures and biomolecules based on low-loss vdW material platforms.

**Materials and Methods**

**Sample fabrication.** Thin flakes of *α*-MoO$_3$ were prepared by mechanical exfoliation of bulk van der Waals crystals. The growth of bulk crystals was described in our previous reports[40,42]. Briefly, commercial MoO$_3$ powders (Alfa Aesar) were evaporated and re-deposited inside a quartz tube through chemical vapour deposition. The as-grown bulk crystals were mechanically exfoliated onto a polydimethylsiloxane thin sheet. Then large and thin *α*-MoO$_3$ flakes with high quality were selected for use. Au disks with different diameters were fabricated on SiO$_2$/Si substrates by electron beam lithography, and Au lines were fabricated via focused ion beam milling. Through deterministic alignment under an optical microscope, the *α*-MoO$_3$ flake was finally transferred onto different Au patterns, forming the imaging device structure.

**Optical measurements.** The infrared nano-imaging measurements were performed using a scattering-type scanning near-field optical microscope (*s*-SNOM, NeaSpec GmbH). In this system, a metallic atomic force microscope tip (NanoWorld ARROW-NCPt Probes), oscillating at a frequency of ~300 kHz and an amplitude of ~70 nm, was illuminated by the *p*-polarized infrared light with tuneable frequencies. The resulting nanoscale hotspot at the apex of the tip interacts with the sample, with the tip acting as both an infrared antenna and a near-field probe. A pseudo-heterodyne interferometer was the used to record the tip-scattered field, and the second or third harmonics of the tip resonant frequency was demodulated to extract the near-field images in this work.

**Numerical simulations.** We performed finite element method simulations utilizing the 'Electromagnetic Waves, Frequency Domain' module of the COMSOL commercial software. A perfectly matched layer was applied to the sides of the simulation to eliminate reflections at the boundaries. A transverse magnetic-polarized plane wave was incident vertically from the top of the sample. The dielectric permittivity of the α-MoO$_3$ flake was taken from references[40,41,53]. In the simulations shown in Fig. 1g-h, the thickness of α-MoO$_3$ was set as 500 nm.